\documentclass[preprint2]{emulateapj-rtx4}

\usepackage{txfonts} \usepackage{epsfig} \usepackage{graphicx}
\usepackage{natbib} 
\begin{document}

\shorttitle{Millimeter multiplicity in DR21(OH)}
\shortauthors{Zapata, et al.}  \title{Millimeter 
multiplicity in DR21(OH): outflows, molecular cores and
envelopes}

\author{Luis A. Zapata\altaffilmark{1}, Laurent
  Loinard\altaffilmark{1}, Y.-N. Su\altaffilmark{2}, Luis F. Rodr\'\i guez\altaffilmark{1,3}, 
  Karl M. Menten\altaffilmark{4}, Nimesh
  Patel\altaffilmark{5}, and R. Galv\'an-Madrid\altaffilmark{1,2,5}}

\altaffiltext{1}{Centro de Radioastronom{\'\i}a y Astrof{\'\i}sica,
  Universidad Nacional Aut\'onoma de M\'exico, Morelia 58090, Mexico}
\altaffiltext{2}{Academia Sinica Institute of Astronomy and
  Astrophysics, P.O.  Box 23-141, Taipei 106, Taiwan}
  \altaffiltext{3}{Astronomy Department, Faculty of Science, King Abdulaziz University, 
P.O. Box 80203, Jeddah 21589, Saudi Arabia}
\altaffiltext{4}{Max-Planck Institut f\"ur Radioastronomie, Auf dem
  H\"ugel 69, 53121 Bonn, Germany}
\altaffiltext{5}{Harvard-Smithsonian Center for Astrophysics, 60
  Garden Street, Cambridge MA 02138, USA}

\begin{abstract}
We present sensitive high angular resolution ($\sim$ 1$"$)
millimeter continuum and line observations from the massive star
forming region DR21(OH) located in the Cygnus X molecular cloud.
Within the well-known dusty MM1-2 molecular cores, we report the
detection of a new cluster of about ten compact continuum
millimeter sources with masses between 5 and 24 M$_\odot$, and sizes of a few thousands
of astronomical units. These objects are likely to be large dusty
envelopes surrounding massive protostars, some of them most
probably driving several of the outflows that emanate from this region.
Additionally, we report the detection of strong millimeter emission of
formaldehyde (H$_2$CO) and methanol (CH$_3$OH) near 218 GHz as well as
compact emission from the typical outflow tracers carbon monoxide  and silicon
monoxide (CO and SiO) toward this massive star-forming region. The
H$_2$CO and CH$_3$OH emission is luminous ($\sim$ 10$^{-4}$
L$_{\odot}$), well resolved, and found along the collimated
methanol maser outflow first identified at centimeter wavelengths and
in the sources SMA6 and SMA7. Our
observations suggest that this maser outflow might be energized by a
millimeter source called SMA4 located in the MM2 dusty core. The
CO and SiO emission traces some other collimated outflows that emanate
from MM1-2 cores, and are not related with the low velocity maser
outflow.  
\end{abstract}


\section{Introduction}

DR21(OH) (also known as W75S) is a well-known high-mass star-forming
region due to its richness of centimeter and millimeter maser emission
from numerous transitions {\it e.g.}  OH, H$_2$O and CH$_3$OH
\citep{araya2009,Fish2005,kurtz2004,kolgan1998,magnun1992,plam1990,Fish2011}.
DR21(OH) is located at a distance of about 2 -- 3 kpc
\citep{Oden1993}, and about 3$'$ (assuming a distance of 2 kpc, this is
equivalent to a physical -projected- size of around 2 pc)  
north of the famous HII region DR 21
in the Cygnus X molecular cloud. In this work, we assume a 
distance of 2 kpc to DR21(OH). However, the exactly value  of the distance is uncertain.

Its total luminosity is estimated to
be of about 5 $\times$ 10$^4$ L$_\odot$ \citep{Har1977}. The region
contains two main dusty condensations (MM1 and MM2) that are warm
($\sim$ 50 K and 30 K; Mangum et al.\ 1992), and very massive (350 and 570 M$_\odot$;
Mangum et al.\ 1992, 1993). DR21(OH) and its surroundings have been
studied in numerous molecular transitions of NH$_3$, CS, $^{12}$CO, and
C$^{18}$O \citep{Padin1989,Magnun1991,magnun1992,Rich1994,Lai2003}.

Multiple molecular outflows have been reported to emanate from the
MM1 and MM2 dusty condensations.  A well collimated east-west bipolar
flow driven from within the MM2 condensation has been reported and
discussed by Plambeck \& Menten (1990), Kogan \& Slysh (1998), and
Kurtz et al.\ (2004). The LSR radial velocity of that flow is nearly
ambient ($+$10 to $-5$ km s$^{-1}$). The LSR radial velocity of DR21(OH) 
is about $-$3.0 km s$^{-1}$, this value was found by \citet{araya2009} 
using the 44 GHz methanol maser line. The detailed distributions of
36 and 44 GHz methanol maser emission toward that flow have been
established by \citet{Fish2011} and \citet{araya2009}, respectively.  
Additionally, \citet{Lai2003} reported high-velocity $^{12}$CO(2-1) outflows with 
v $\geq$ 25 km s$^{-1}$ relative to the systemic velocity powered by
MM1-2. A CO bipolar outflow expels material to the northwest
(blueshifted) and southeast (redshifted) and is originating from MM
2. A second bipolar outflow emanates from MM1-2 with its
blueshifted lobe towards the southwest while its redshifted one is to the
northeast. \citet{Lai2003} also suggested the possibility of
having a single EW bipolar outflow with a conelike morphology, with
the CO lobes tracing the limb-brightened region of the outflow and
emanating from MM1-2.  \cite{Rich1994} reported high velocity wings in
CS(J=5-4) towards DR 21(OH), extending over 80 km
s$^{-1}$, probably produced by a young and compact outflow.
 
In this paper, we present high angular resolution ($\sim$ 1$''$ or 2000 AU)
millimeter and submillimeter continuum and line observations of the
region DR21(OH) made with the Submillimeter Array.  In Section 2, we
discuss the observations undertook in this study. In Section 3, we
present and discuss the data, and in Section 4 we give the main
conclusions of this study.
 
  \begin{figure*}[ht]
\begin{center}\bigskip
\includegraphics[scale=0.2]{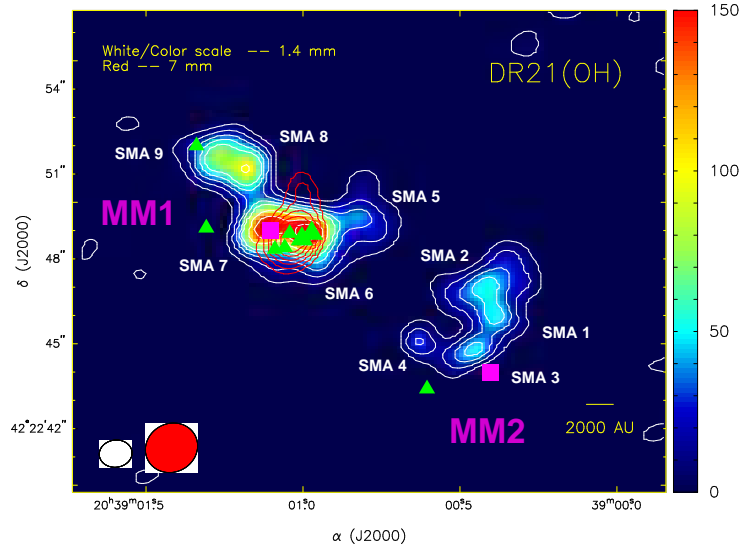}
\caption{\scriptsize SMA 1.4 mm  continuum (color and white contour image) of the DR21(OH) region
  overlaid with the 7 mm continuum
  emission from Zapata et al.\ 2009 (red contours). The color-scale bar on the
  right indicates the 1.4 mm continuum emission in mJy
  beam$^{-1}$. The white contours are 2, 4, 6, 8, 10, 15, 20, 25, 30,
  35, and 40 times 4 mJy beam$^{-1}$, the rms noise of the 1.4 mm
  image. The red contours are -5, 5, 6, 7, 8, 9, and 10 times 0.9
  mJy beam$^{-1}$, the rms noise of the 7 mm image. The half-power
  contours of the synthesized beams for the different wavelengths are
  shown in the bottom-left corner of the image. For the 1.4 mm
  observations the beam is $1\rlap.{''}16$ $\times$ $0\rlap.{''}96$ with a
  P.A. = $-82.67^\circ$ and for the 7 mm observations is
  $1\rlap.{''}88$ $\times$ $1\rlap.{''}71$ with a P.A. =
  $-61.12^\circ$. The green triangles mark the positions of the radio
  sources reported by \citet{araya2009}. The purple squares represent
  the positions of the 2.7 mm sources (with angular sizes of about
  10$''$)
  reported by \citet{Magnun1991}.}
\label{fig1}
\end{center}
\end{figure*}
 
\section{Observations}

The millimeter ($\nu$ $\simeq$ 217 -- 230 GHz or $\lambda$ $\simeq$ 1.4 -- 1.3 mm) 
observations of DR21(OH) were
obtained with the Submillimeter Array\footnote{The Submillimeter
Array (SMA) is a joint project between the Smithsonian Astrophysical
Observatory and the Academia Sinica Institute of Astronomy and
Astrophysics, and is funded by the Smithsonian Institution and the
Academia Sinica.} (SMA) in 2006 May 23 and 2007 August 26. The SMA at those epochs
time was in its extended configuration, which included 28 independent
baselines ranging in projected length from 18 to 162 k$\lambda$. The
phase reference center was $\alpha(J2000.0)$ =
20$^h$39$^m$00.0$^s$, $\delta(J2000.0)$ =
$+$42$^\circ$22$'$48.0$''$. The central frequency of the Lower
Sideband (LSB) was 217.1049 GHz, while the Upper Sideband (USB)
central frequency was 227.1049 GHz for the night in May 2006. The
frequency was centered for the night of 2007 August at 230.570 GHz 
in the LSB, while the USB was centered at 220.570 GHz.

The full bandwidth of the SMA correlator was 4 GHz (2 GHz in each
band).  For the night in May 2006, the SMA digital correlator was
configured in 24 spectral windows (``chunks'') of 104 MHz each, with
128 channels distributed over each spectral window, providing a
resolution of 0.813 MHz ($\sim$ 1 km s$^{-1}$) per channel.  For the
night of 2007 August  256 channels distributed over each spectral
window were used, providing a resolution of 0.41 MHz ($\sim$ 0.5 km s$^{-1}$)
per channel. In both observation the linear polarization receivers were used.
The primary beam at 230 GHz has a FWHM diameter of about 50$''$.
The greatest angular size source that can be imaged on these observations
is approximately 15$''$.  The astrometric errors of these observations are 
less than 1$''$.

The zenith opacity ($\tau_{230 GHz}$), measured with the NRAO tipping
radiometer located at the Caltech Submillimeter Observatory, was
between 0.09 and 0.25, indicating good weather conditions during the
observations. Observations of Callisto and Uranus provided the
absolute scale for the flux density calibration. In the night of 2007 August,
we observed Uranus, while for the night of May 2006 Callisto was used. 

The gain calibrators were the quasars 2015+371 and BL Lac 
(with flux densities of 1.5 and 2.5 Jy, respectively).  
The uncertainty in the flux scale is estimated to be 15 -- 20$\%$, 
based on the SMA monitoring of quasars. 
For the May 2006 observations, we integrated on-source for a total of 
approximately 2 hr and for the night of 2007 August about of 3 hrs.
Further technical descriptions of the SMA and its
calibration schemes can be found in \citet{Hoetal2004}.

The data were calibrated using the IDL superset MIR, originally
developed for the Owens Valley Radio Observatory
\citep{Scovilleetal1993} and adapted for the SMA.\footnote{The MIR-IDL
  cookbook by C. Qi can be found at
  http://cfa-www.harvard.edu/$\sim$cqi/mircook.html} The calibrated
data were imaged and analyzed in the standard manner using the MIRIAD
and AIPS packages. 

To obtain the continuum map, we only use the data from the night of May 2006
due to the phase stability was slightly better.
The resulting image rms noise of line and continuum
images were around 20 mJy beam$^{-1}$ for each velocity channel and 5
mJy beam$^{-1}$ for a 2 GHz total bandwidth, respectively, at an
angular resolution of $1\rlap.{''}16$ $\times$ $0\rlap.{''}96$ with a
P.A. = $-82.67^\circ$.  The resulting continuum map was
self-calibrated in phase.  We use the LSB of the 2006 run to
reconstruct the line-free continuum image since the USB is more affected by 
line emission from different molecular species.

 \begin{table*}
\begin{center}
\scriptsize
\caption{Parameters of the 1.4 mm continuum sources in DR21(OH)}
\begin{tabular}{lcccccc}
\tableline 
          & \multicolumn{2}{c}{Position$^a$} &  \\
\cline{2-3} Source & $\alpha$(J2000) & $\delta$(J2000) & Flux Density  & Deconvolved Angular Size$^b$ & Mass\\ 
                  & 20 39 & 42 22 &  [mJy]   &  & [M$_\odot$]\\ 
\tableline 
SMA 1 & 00.400 & 46.60 & 122$\pm$24    & $3\rlap.{''}6 \pm 0\rlap.{''}3 \times 1\rlap.{''}3 \pm 0\rlap.{''}2;~ +1^\circ \pm 10^\circ$    & 8\\ 
SMA 2 & 00.406 & 46.78 & 185$\pm$20    & $2\rlap.{''}1 \pm 0\rlap.{''}2 \times 1\rlap.{''}5 \pm 0\rlap.{''}2;~ +137^\circ \pm 32^\circ$  & 12 \\ 
SMA 3 & 00.453 & 44.86 & 80$\pm$9      & $\leq$ $2\rlap.{''}0$                                                                         & 6 \\ 
SMA 4 & 00.598 & 44.94 & 49$\pm$9      & $\leq$ $2\rlap.{''}0$                                                                         & 5\\ 
SMA 5 & 01.010 & 48.72 & 44$\pm$8      & $4\rlap.{''}5 \pm 0\rlap.{''}2 \times 1\rlap.{''}4 \pm 0\rlap.{''}2;~ +104^\circ \pm 2^\circ$   & 5 \\ 
SMA 6 & 01.002 & 48.93 & 346$\pm$12    &$1\rlap.{''}90 \pm 0\rlap.{''}09 \times 1\rlap.{''}00 \pm 0\rlap.{''}09;~ +77^\circ \pm 3^\circ$ & 23 \\ 
SMA 7 & 01.079 & 49.06 & 357$\pm$13    &$1\rlap.{''}8 \pm 0\rlap.{''}1 \times 1\rlap.{''}13 \pm 0\rlap.{''}1;~ +88^\circ \pm 4^\circ$    & 24 \\ 
SMA 8 & 01.192 & 51.26 & 201$\pm$13    &$1\rlap.{''}9 \pm 0\rlap.{''}1 \times 0\rlap.{''}8 \pm 0\rlap.{''}1;~ +30^\circ \pm 5^\circ$     & 14 \\ 
SMA 9 & 01.231 & 51.43 & 217$\pm$14    &$1\rlap.{''}9 \pm 0\rlap.{''}1 \times 1\rlap.{''}1 \pm 0\rlap.{''}1;~ +72^\circ \pm 7^\circ$     & 14\\ 
\tableline
\end{tabular}
\tablenotetext{a}{Units of right ascension are hours, minutes, and
  seconds and units of declination are degrees, arcminutes, and
  arcseconds.}
\tablenotetext{b}{Major axis $\times$ minor axis; position angle of
  major axis. The values were obtained using the task JMFIT
  of AIPS.} 
\end{center}
\end{table*}

\section{Results and discussion}

\subsection{Continuum emission}

In Figure 1, we show a color and contour map of the 1.4 mm continuum emission
detected by the SMA and the VLA towards DR21(OH).  
We resolved the strong 2.7 millimeter sources MM1 and MM2 reported by \citet{Magnun1991,
  magnun1992} into a cluster of nine compact sources. 
Here, the term ``compact sources'' is with respect to the size of the extended 
sources reported by \citet{Magnun1991,magnun1992}. 
Five of these sources are associated 
with MM1 (SMA5-9) and four with MM2
(SMA1-4). We give their positions and
total flux densities in Table 1. Additionally, in Figure 1, we have
overlaid a 7 mm contour continuum map obtained from \citet{Zapata2009}
and the positions of the centimeter compact sources reported by
\citet{araya2009}.  The 7 mm continuum emission peaks at the position
of the centimeter sources (NW+SE+R2+R3+R4) reported by
\citet{araya2009} and is not coincident with any SMA 1.4 mm
continuum sources. Only SMA6 has a clear counterpart at centimeter
wavelengths (MM1-NW and MM1-SE). The 1.4 mm sources are well resolved 
at these wavelengths and show sizes of a few thousands of Astronomical Units 
at an assumed distance of 2 kpc.

Assuming a very steep spectral index of $\alpha$=3.5 (S$_\nu$ $\propto$ $\nu^\alpha$) 
for all the millimeter sources, which is consistent with
optically thin dust emission from dusty envelopes or disks,
we can estimate the masses of the 1.4 mm sources. 
These steep spectral indices have been observed in many star forming regions and
are associated with very young stellar objects \citep[see for example][]{hunter2006,rod2007,roberto2010}. 

Following \citet{beck1990}, we adopt a value for the dust mass opacity of
$\kappa_{\nu}$ = 0.1 ($\nu$/1000 GHz)$^\beta$ cm$^2$ g$^{-1}$, where $\nu$ is the frequency and here 
$\beta$ = $\alpha$-2 = 1.5. Thus, at this wavelength, we obtain 
$\kappa_{1.4 mm}$ = 0.01 cm$^2$ g$^{-1}$.  Assuming optically thin, isothermal dust emission and a
gas-to-dust ratio of 100, the total mass of the 1.4 mm sources is
given by

$$\Biggl[{{M_{gas}} \over {M_\odot}}\Biggr] = 1.6 \times 10^{-6} \Biggl[{{S_\nu}
    \over {Jy}}\Biggr] \Biggl[{{T} \over {K}}\Biggr]^{-1} \Biggl[{{D}
    \over {pc}}\Biggr]^{2}\Biggl[{{\nu} \over {1000~GHz}}\Biggr]^{-(2+\beta)},$$
    
\noindent
where S$_\nu$ is the flux density, T is the dust temperature and D is
the distance to the source.  Assuming a temperature of T = 20 K for
all the millimeter sources, we derive masses on the
range of 5 -- 24 solar masses for the sources (see Table 1). 
The gas-to-dust ratio of 100 might not be the most adequate to use 
for protostellar sources since dust settling to the midplane of 
the disk and erosion of the circumstellar envelope by photodissociation
may decrease the gas-to-dust ratio \citep{troop2005}. The sources
SMA6 and SMA7 seem to show hot core activity and their temperatures
could be higher, so the estimation of the mass for these sources
might be overestimated. For the rest of the sources a temperature 
of 20 K seems to be more adequate because they do not show hot core activity.

Since SMA6 is associated with free-free emission \citep{araya2009}, the emission
at 1.4 mm may be is contaminated with this type of emission. 
However, this contamination seems to be
almost negligible at these wavelengths due to the relatively flat spectral 
index ($\alpha$ = 0.6) obtained at centimeter wavelengths for this source \citep{araya2009}. 

The values of the masses obtained here (Table 1) have uncertainties of at least  2 or 
larger due to the error in the determination of the distance to DR21(OH),
the estimation of temperatures of the millimeter sources, and
the error in the dust mass opacity coefficient at this wavelength.
 
Similar dust mass values have been recently found for the gas structures associated with the
massive protostars in the young clusters W33A and NGC6334N(I)
\citep{roberto2010,hunter2006}.

\begin{figure*}[ht]
\begin{center}
\includegraphics[scale=0.7]{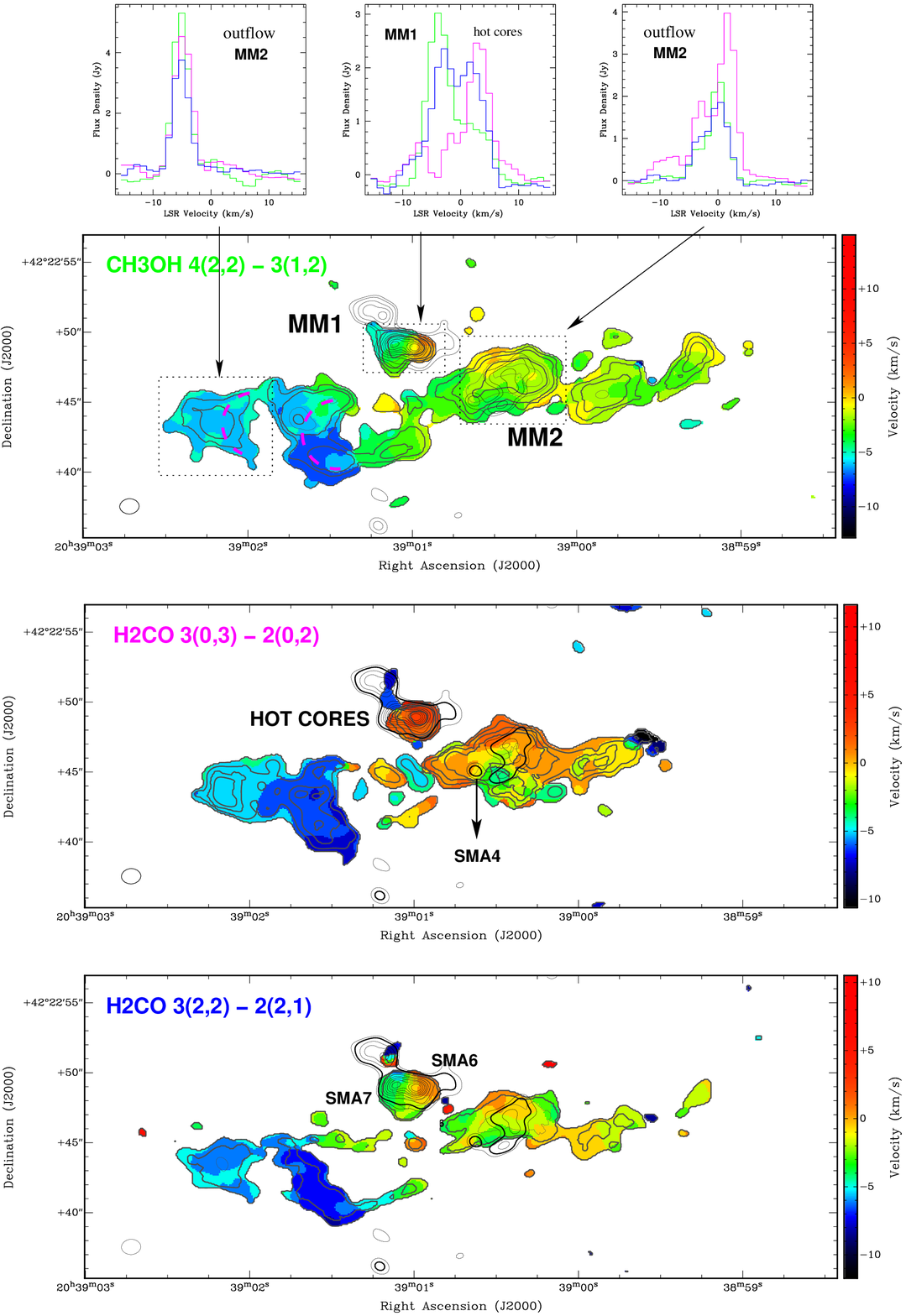}
\caption{\scriptsize Integrated intensity of the weighted velocity color maps
  of the CH$_3$OH[4(2,2)-3(1,2)-E] (upper panel), H$_2$CO[3(0,3)-
    2(0,2)] (middle panel), and H$_2$CO[3(2,2)- 2(2,1)] (lower panel),
  emission from the DR21(OH) region overlaid in contours with the SMA
  1.4 mm continuum emission (black thick line) and the integrated
  intensity emission of the specific molecule (grey thin line) in
  every panel. The black contours are from 30\% to 94\% with steps of
  8\% of the peak of the line emission; the peak 1.4 mm emission is
  150 mJy Beam$^{-1}$. The grey contours are from 5\% to 85\% with
  steps of 10\% of the peak of the line emission. The color-scale bars
  on the right indicate the LSR velocities in km s$^{-1}$. The three
  spectra shown on the top of the panels were obtained from different
  positions across the outflow and the molecular cores as
  indicated. The colors of the spectra indicate the transition
  (green=CH$_3$OH[4(2,2)-3(1,2)-E], pink=H$_2$CO[3(0,3)-2(0,2)], and
  blue=H$_2$CO[3(2,2)- 2(2,1)]). The synthesized beam of the CH$_3$OH
  image is shown in the bottom left corner of the image. The pink
  dashed arcs in the top panel indicate the same "arc'' morphology found 
  in the outflow in the methanol masers at centimeter wavelengths \citep{araya2009}.  }
\label{fig2}
\end{center}
\end{figure*}

\begin{table*}[ht]
\scriptsize 
\caption{Observational and physical parameters of the millimeter lines}
\begin{center}
\begin{tabular}{lccccccc}
\hline \hline & Rest frequency$^a$  & E$_{lower}$ & Range of Velocities & Linewidth$^b$ & LSR 
Velocity$^b$ & Intensity Peak \\ 
Lines & [GHz] & [K] & [km s$^{-1}$] & [km s$^{-1}$] & [km s$^{-1}$] & [mJy Beam$^{-1}$] \\  \hline 
H$_2$CO[3(0,3)-2(0,2)]    & 218.22219  & 10.4   & $-$10,$+$10 &  14 & $-$3 & 80\\ 
H$_2$CO[3(2,2)-2(2,1)]    & 218.47563  & 57.6   & $-$10,$+$11 &  13 & $-$3 & 60\\ 
CH$_3$OH[4(2,2)-3(1,2)-E] & 218.44005  & 35.0   & $-$11,$+$10 &  13 & $-$3 & 78\\
SiO(5-4)                  & 217.10490  & 20.8   & $-$11,$+$7  &  10 & $-$3 & 26\\
$^{12}$CO($J=2-1$)         & 230.53801  & 05.3   & $-$40,$+$30 &  40 & $-$5 & 170 \\   \hline
\end{tabular}
\tablenotetext{a}{The rest frequencies were obtained from the the JPL Molecule Catalog: 
http://spec.jpl.nasa.gov/ftp/pub/catalog/catform.html.}
\tablenotetext{b}{The line-width and LSR velocity were obtained fitting a Gaussian profile to the spectra.}
\end{center}
\end{table*}

\subsection{Millimeter line emission}

Five strong spectral lines were detected in the LSB (2006 May) and in
the USB (2007 August) of the observations, corresponding to
the H$_2$CO[3(0,3)-2(0,2)], H$_2$CO[3(2,2)-2(2,1)], CH$_3$OH[4(2,2)-3(1,2)-E], 
SiO(5-4), and $^{12}$CO(2-1) transitions (see Table 2).

\subsubsection{H$_2$CO and  CH$_3$OH}

Figure 2 shows maps of the H$_2$CO[3(0,3)-2(0,2)],
H$_2$CO[3(2,2)-2(2,1)], and CH$_3$OH[4(2,2)-3(1,2)-E] integrated
intensity (moment 0) and intensity-weighted velocity (moment 1),
overlaid with the 1.4 mm continuum emission obtained in our SMA observations.  
These maps reveal strong line
millimeter molecular emission arising from the east-west outflow and
from the two compact continuum sources SMA6 and SMA7, first reported
here (see Section 3.2). The three lines show comparable integrated
flux densities.  The radial velocities covered by the outflow are from
$-$8 to $+$5 km s$^{-1}$, whereas those corresponding to the molecular core
sources range from $-$11 to $+$11 km s$^{-1}$.

The emission appears to be concentrated in compact bow-shock
structures within the outflow where the line profiles are narrow (a few
km s$^{-1}$) and intense ($\sim$ 3 - 5 Jy). Overall, the methanol and
formaldehyde millimeter emission reported here follow a morphology
very similar to that seen in the 44 GHz methanol maser line
\citep{araya2009}. This combination of properties suggest
that the formaldehyde and methanol spectral lines reported here might
correspond to maser transitions. However, the low brightness
temperatures of all lines (T$_B$ $\sim$ 280, 200, 70 K for the
CH$_3$OH, H$_2$CO[3(2,2)-2(2,1)], and H$_2$CO[3(0,3)-2(0,2)],
respectively, and assuming the emission extends over all the beam) 
suggest thermal emission. These low brightness
temperatures, however, are also seen in most of the 44 GHz methanol
masers at centimeter wavelengths \citep{araya2009}. One possibility is
that the millimeter line spots detected here are probably much more
compact and not resolved with our present angular resolution ($\sim$
1$''$) resulting thus in true maser emission. More observations are
needed to confirm if these millimeter lines are masing.  The flux of
all three lines integrated over the entire outflow is 80 to 150 Jy km
s$^{-1}$, corresponding to isotropic luminosities of about 10$^{-4}$
L$_\odot$. This is comparable to the luminosity of other typical maser
lines in star-forming regions \citep{Zapata2009}.

\begin{figure*}[ht]
\begin{center}
\includegraphics[scale=0.43]{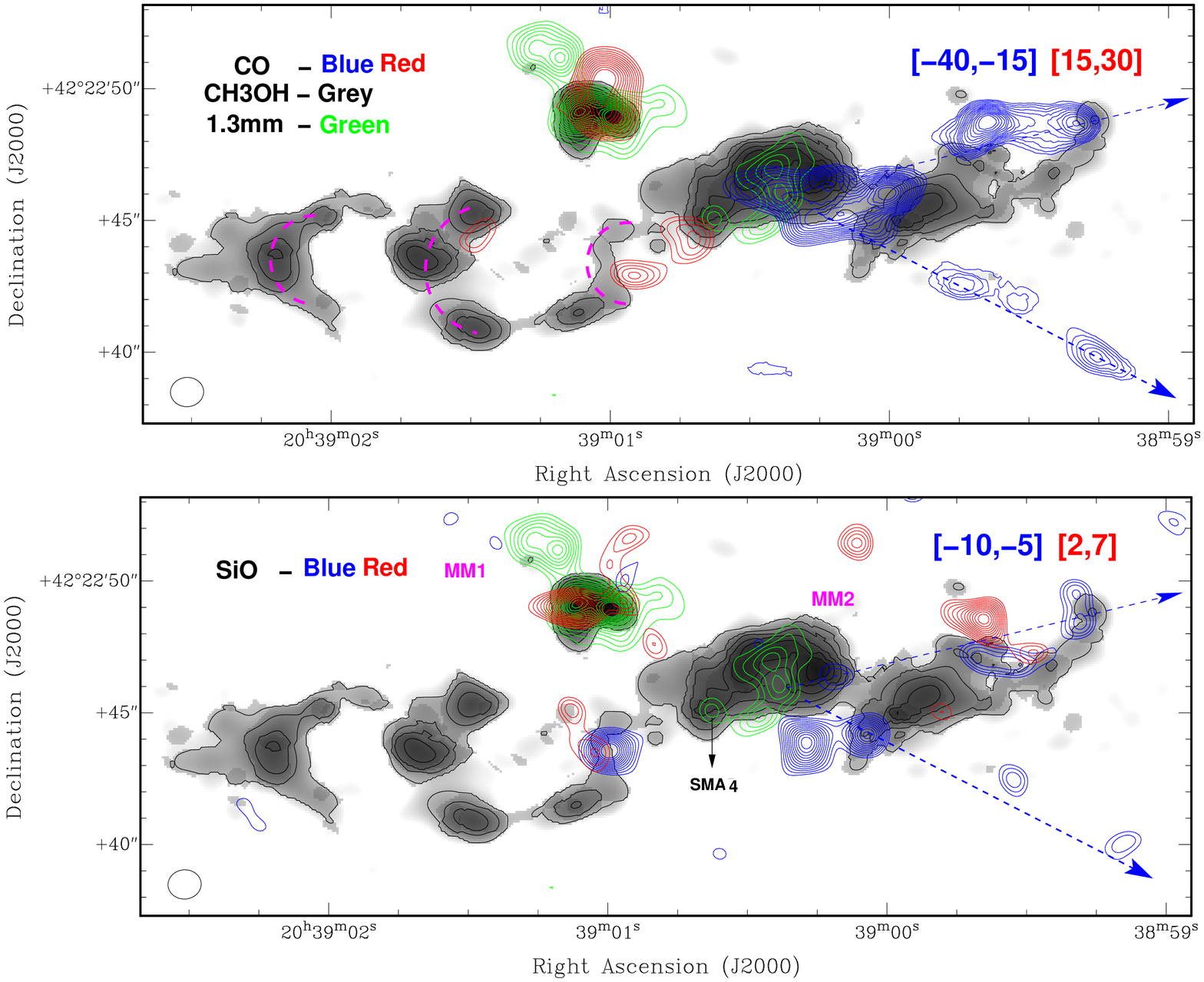}
\caption{ \scriptsize {\it Upper:} Integrated intensity 
  grey-scale map of the CH$_3$OH[4(2,2)-3(1,2)-E] emission from the
  region DR21(OH) overlaid in contours with the 1.4 mm continuum
  emission (green) and the integrated intensity emission of the
  $^{12}$CO(2-1) (blue/red). The integrated velocity range for the
  blueshifted $^{12}$CO(2-1) gas is from $-$40 to $-$15 km s$^{-1}$
  and that for the redshifted gas is from $+$15 to $+$30 km
  s$^{-1}$. These values are indicated in the top right corner of the
  image.  The integrated velocity range for the CH$_3$OH gas is from
  -10 to 10 km s$^{-1}$. The synthesized beam of the CH$_3$OH image is
  shown in the bottom left corner of the image. The $^{12}$CO(2-1)
  blue/red contours are from 40\% to 90\% with steps of 5\% of the
  peak of the line emission. The peak of the $^{12}$CO(2-1) emission
  is about 12 Jy Beam$^{-1}$ km s$^{-1}$.  The black contours are from
  10\% to 90\% with steps of 10\% of the peak of the line emission;
  the peak CH$_3$OH emission is 5.5 Jy beam$^{-1}$ km s$^{-1}$. {\it
    Lower:} Same as the upper image, but with the SiO(5-4) emission
  instead of the $^{12}$CO(2-1). The integrated velocity range for the
  blueshifted SiO(5-4) gas is from $-$10 to $-$5 km s$^{-1}$, and that
  for the redshifted gas is from $+$2 to $+$7 km s$^{-1}$. The
  SiO(5-4) blue/red contours are from 30\% to 90\% with steps of 4\%
  of the peak of the line emission. The pink
  dashed arcs in the top panel indicate the same "arc'' morphology found  
  in the methanol masers at centimeter wavelengths \citep{araya2009}. 
  The arrows mark the two eastern blue-shifted outflows emanating from the 
  zone. }
\label{fig3}
\end{center}
\end{figure*}

This molecular emission, with low radial velocities that suggest motion near
the plane of the sky, is reminiscent of water maser emission tracing outflows
that are known to be within a few degrees from the plane of the sky
\citep{Claussenetal1998,Des2009}.

At the center of symmetry of the outflow lies a continuum source
(labelled SMA4 on Figure 2, middle panel). 
The source likely traces the envelope of
a high- or intermediate- mass young star (Table 1) and is part of the larger dusty core
MM2. The``blue'' infrared source proposed by \citet{araya2009} as a
possible candidate for the exciting source of the east-west flow is a
bit offset (2$''$) from SMA4 and the center of symmetry of the
outflow. The symmetry center of the outflow is approximately where
the systemic cloud velocity resides, that is, in the middle of the outflow.
The blueshifted velocities are found toward the east while the redshifted
velocities toward the west.   
Clearly, more observations will be required to discriminate
firmly the true powering source of the outflow, and to examine the
relation between SMA4 and the blue infrared source identified by
\citet{araya2009}.

In contrast to the narrow spectra found within the outflow, the
methanol and formaldehyde line profiles associated with the sources
SMA6 and SMA7 are broad ($\sim$ 15 km s$^{-1}$) and with the morphology being much more compact. 
This emission may trace hot molecular core emission associated with these dusty
objects. However, a more complete molecular line analysis is required to firmly
confirm this hypothesis. Furthermore, both lines (CH$_3$OH and H$_2$CO) show a clear E-W velocity
gradient of a few kilometers within the molecular core, probably
suggesting that SMA6 and SMA7 are at slightly different systemic
velocities or maybe that emission is tracing a molecular compact outflow.

\subsubsection{$^{12}$CO and SiO}

Together with the H$_2$CO and  CH$_3$OH observations, we obtained observations 
of the classical outflow tracers $^{12}$CO(2-1)
and SiO(5-4) to study in-deep the methanol maser outflow, however,
such emission was not detected at all toward the methanol outflow.  
We instead found some other compact high/low velocity outflows within the region
emanating from MM1 or MM2 (see Figure 3). Some of these outflows were already
reported with less angular resolution ($\sim$ 4$"$) by \citet{Lai2003}
in $^{12}$CO(2-1).
   
The observations of the $^{12}$CO(2-1) revealed two collimated
outflows emanating from MM2, one high-velocity bipolar with its
redshifted side ($+$30 to $+$15 km s$^{-1}$) in the east and with its
blueshifted side ($-$65 to $-$15 km s$^{-1}$) toward the west, and a
second monopolar outflow with its blueshifted emission ($-$40 to $-$15 km
s$^{-1}$) toward the southwest.  None of these outflows are
associated with the low-velocity ($-$10 to $+7$ km s$^{-1}$) E-W
methanol maser bipolar outflow.  This might be explained as a result
of removing the $^{12}$CO emission from velocities close to ambient in order
to make the map presented in Figure 3, however, as we will see below,
even the SiO, an outflow tracer that is supposedly weak at
ambient velocities, is not present at the position of the maser
outflow. Furthermore, one would think that the EW carbon monoxide
outflow could be the counterpart of EW maser methanol outflow,
however, the blueshifted and redshifted sides of both outflows are
found in contrary positions, see Figures 2 and 3.

There is one more very compact north-south $^{12}$CO(2-1) outflow
emanating from the MM1 core with its redshifted side to the north,
powered maybe by SMA6 or SMA7. Its blueshifted side is not detected.

Our $^{12}$CO(2-1) map does not favor the idea of having a single
bipolar east-west outflow with a conelike morphology, with the CO
lobes tracing the limb-brightened region of the outflow as suggested
by \citet{Lai2003}. Our results instead confirm the presence of
multiple compact outflows with different orientations emanating within the
MM1-2 cores.

The SiO(5-4) on the other hand, shows a more clumpy structure
over the whole region.  The radial velocities displayed by this
molecule are very similar to the ones of the methanol maser bipolar
outflow ($-$10 to $+7$ km s$^{-1}$), however the SiO emission is not
arising from this outflow (see Figure 3).  There is instead faint
blueshifted emission clearly associated with the east-west and
southeast-northwest $^{12}$CO(2-1) outflows.  Towards the MM1 core,
the SiO traces some other different compact outflows than the
north-south outflow revealed by the $^{12}$CO(2-1) emission.  However,
their orientations are not clear from the present observations.  Some
of these molecular outflows could be are powered by the thermal jets
(MM1-NW and MM1-SE) reported toward this position by
\citet{araya2009}.

\section{Summary}

We have reported the detection of a new cluster of about ten compact
millimeter sources with masses in a range of 5 to 24 M$_\odot$ at
the center of DR21(OH). These sources are likely to be large dusty
envelopes surrounding high- or intermediate- mass protostars, and some of them most
probably drive multiple outflows that emanate from this region.

We also reported for the first time the detection of strong millimeter
emission of formaldehyde (H$_2$CO) as well as the methanol (CH$_3$OH)
at around 218 GHz towards DR21(OH).  The line emission is detected within the
east-west flow driven by DR21(OH) and the sources SMA6 and SMA7, 
and is well coincident with methanol centimeter (36 and 44 GHz) maser 
emission previously reported. 

The SiO and $^{12}$CO emission revealed a group of compact outflows emerging
from the cluster of young stars present in DR21(OH).  We do
not find $^{12}$CO high velocity emission neither low velocity SiO emission coincident with 
the methanol maser outflow.

\acknowledgments 

We would like to thank the anonymous referee 
for valuable suggestions that improve this paper.
L.A.Z., L.L., and L.F.R.\ acknowledge the financial
support from DGAPA, UNAM, and CONACyT, M\'exico. 
L.L.\ is indebted to the Guggenheim Foundation for financial support.

\end{document}